\documentclass[dvips,12pt]{article}
\usepackage{hyperref}
\usepackage{amsfonts}
\usepackage{amsmath, amssymb}
\usepackage{graphicx}
\usepackage{psfrag}
\usepackage{youngtab}
\Yboxdim{5pt}

\newcommand{\ra}{\rightarrow}

\newcommand{\be}{\begin{equation}}
\newcommand{\ee}{\end{equation}}
\newcommand{\ba}{\begin{eqnarray}}
\newcommand{\ea}{\end{eqnarray}}
\newcommand{\bi}{\begin{itemize}}
\newcommand{\ei}{\end{itemize}}

\newcommand{\Tr}{{\rm Tr}}

\newcommand{\R}{{\rm R}}
\newcommand{\C}{{\bf C}}
\newcommand{\p}{\partial}

\newcommand{\CP}{{\bf P}}

\newcommand{\Ncal}{{\mathcal N}}

\newcommand{\Fcal}{{\mathcal F}}

\newcommand{\Jcal}{{\mathcal J}}
\newcommand{\Acal}{{\mathcal A}}

\newcommand{\Kahler}{K\"{a}hler }

\newcommand{\nn}{\nonumber}
\newcommand{\mo}{{-1}} 

\newcommand{\f}{\frac}
\newcommand{\half}{\frac{1}{2}}
\newcommand{\oo}{\frac{1}}

\def\Dslash{\,\,{\raise.15ex\hbox{/}\mkern-12mu D}}
\def\Dbarslash{\,\,{\raise.15ex\hbox{/}\mkern-12mu {\bar D}}}
\def\delslash{\,\,{\raise.15ex\hbox{/}\mkern-9mu \partial}}
\def\delbarslash{\,\,{\raise.15ex\hbox{/}\mkern-9mu {\bar\partial}}}
\def\pslash{\,\,{\raise.15ex\hbox{/}\mkern-9mu p}}
\def\calDslash{\,\,{\raise.15ex\hbox{/}\mkern-12mu {\cal D}}}

\renewcommand{\bar}{\overline}

\begin{document}
\baselineskip=15.5pt
\renewcommand{\theequation}{\arabic{section}.\arabic{equation}}
\pagestyle{plain} \setcounter{page}{1}
\bibliographystyle{utcaps}
\begin{titlepage}

\leftline{\tt hep-th/0612190}

\vskip -.8cm

\rightline{\small{\tt NSF-KITP-06-121}}

\begin{center}

\vskip 3 cm

{\Large {\bf  Wilson Loops, Geometric Transitions\\
\vskip 0.5cm and
Bubbling Calabi-Yau's}}

\vskip 1.5cm
{\large Jaume Gomis\footnote{jgomis at perimeterinstitute.ca} and Takuya Okuda\footnote{takuya at kitp.ucsb.edu}}

\vskip 0.8cm

Perimeter Institute for Theoretical Physics

Waterloo, Ontario N2L 2Y5, Canada$^{1}$

and

Kavli Institute for Theoretical Physics

University of California,
Santa Barbara

CA 93106, USA${}^2$

\vskip 1.5cm

{\bf Abstract}

\end{center}

Motivated by recent developments in the AdS/CFT correspondence,
we provide several alternative bulk descriptions of an arbitrary Wilson loop operator in Chern-Simons
theory.
Wilson loop operators in Chern-Simons theory can be
given a  description in terms of a
 configuration of branes or alternatively anti-branes in the resolved conifold geometry.
The representation of the Wilson loop is encoded in the holonomy of the gauge field
living on the dual brane configuration.
By letting the branes undergo a new type of geometric transition,
we argue that each Wilson loop operator can also be described by a bubbling Calabi-Yau geometry,
whose topology encodes the representation of the Wilson loop.
These Calabi-Yau manifolds provide  a novel representation of knot invariants.
For the unknot we confirm these identifications to all orders in the genus expansion.

\end{titlepage}

\newpage

\tableofcontents
\section{Introduction, summary  and conclusion}

The study of quantum gravity using a holographic   formulation requires
understanding how the bulk gravitational physics is described by the boundary gauge theory. Given the ubiquitous nature of  Wilson loop
operators in gauge theory, it is  of interest to understand how   Wilson loop operators in a gauge theory participating in the holographic formulation of a bulk theory are encoded in the corresponding bulk description. Since gauge theories can be formulated using Wilson loops as the basic variables, an improved understanding of holography may arise from identifying these variables in the bulk theory.

This program has recently been completed in \cite{Gomis:2006sb} for
all half-BPS Wilson loop operators in ${\cal N}=4$ SYM (for previous
work see \cite{Drukker:2005kx,Yamaguchi:2006te}). This extends the
bulk description of a Wilson loop in the fundamental representation
in terms of a string world-sheet \cite{Rey:1998ik,Maldacena:1998im}
to all other representations.  It was found that for each Wilson
loop, there is a bulk description either in terms of a configuration
of D5-branes or alternatively in terms of a configuration of
D3-branes in  AdS$_5\times$S$^5$ (see \cite{Gomis:2006sb} for
details). The equations determining the supergravity background
produced by these D-branes were found in
\cite{Yamaguchi:2006te,Lunin:2006xr} (see also \cite{Gomis:2006cu}
for the closely related equations for half-BPS domain wall
operators), generalizing the supergravity solutions of  LLM
\cite{Lin:2004nb} dual to  half-BPS local operators to half-BPS
Wilson loop operators. All these asymptotically AdS$_5\times$ S$^5$
solutions encode in their nontrivial topology the information about
the dual operator. For other work see e.g.
\cite{Yamaguchi:2006tq,Hartnoll:2006hr,Okuyama:2006jc,Hartnoll:2006is,Giombi:2006de,Tai:2006bt,Gomis:2006im}.

A very interesting and calculable example of a holographic
correspondence is the one found by Gopakumar and Vafa
\cite{Gopakumar:1998ki} in   topological string theory\footnote{See
\cite{Neitzke:2004ni} for an overview of results and the book
\cite{Marino:2005sj} by Mari\~no for a comprehensive introduction.}.
The duality states that the A-model topological string theory on the
deformed conifold $T^*{\rm S}^3$ in the presence of $N$ D-branes
wrapping  ${\rm S}^3$ is dual to the A-model topological  string
theory on the resolved conifold geometry, with the complexified
\Kahler modulus given by $t=g_s N$, where $g_s$ is the string
coupling constant on both sides of the correspondence. As shown by
Witten \cite{Witten:1992fb}, the  physics on the deformed conifold
is described  by $U(N)$ Chern-Simons theory on ${\rm S}^3$.

The goal of this paper is to give the bulk  description of Wilson
loop operators in Chern-Simons theory\footnote{As suggested in
\cite{Gopakumar:1998ki}, the Wilson loops can be described by string
world-sheets in the resolved conifold.  We elaborate on this
description in Appendix \ref{F1}.}. As we shall see, a picture
closely related to the one for ${\cal N}=4$ SYM emerges, with the
added benefit that we have control over all quantum corrections for
the case of Chern-Simons theory.  We find that Wilson loops in
Chern-Simons can be described either in terms of a configuration of
D-branes or a configuration of anti-branes in the resolved conifold
geometry. Moreover, by letting the branes undergo a geometric
transition, we find that Wilson loops can be described in terms of
bubbling Calabi-Yau geometries, with rich topology.

We first identify the holographic description of Wilson loop operators
in Chern-Simons theory on ${\rm S}^3$ -- labeled by a knot in ${\rm S}^3$ and a representation of $U(N)$ --
in terms of branes in the resolved conifold geometry.
Just like in \cite{Gomis:2006sb}, we find that  there are two different brane configurations
corresponding to each Wilson loop operator.

The information about the knot $\alpha\subset {\rm S}^3$ is encoded in the choice of a Lagrangian submanifold $L$
in the resolved conifold geometry, as explained by Ooguri and Vafa
 in \cite{Ooguri:1999bv} (see also e.g. \cite{Labastida:2000zp,Labastida:2000yw,Marino:2001re}).
$L$ ends on the knot $\alpha$ on the ${\rm S}^3$ at asymptotic infinity of the resolved conifold geometry,
where the holographic dual Chern-Simons theory lives\footnote{
Taubes \cite{Taubes:2001wk} proposed a procedure to construct the Lagrangian submanifolds
corresponding to the knots that are invariant under the antipodal map of $S^3$.
The construction of the Lagrangians corresponding to arbitrary knots was achieved by Koshkin \cite{Koshkin}.
}.

\begin{figure}[htbp]
 \begin{center}
\psfrag{1}{$R_1$}
\psfrag{2}{$R_2$}
\psfrag{P}{$R_P$}
\psfrag{T1}{$R^T_1$}
\psfrag{T2}{$R^T_2$}
\psfrag{TM}{$R^T_M$}
\includegraphics[width=40mm]{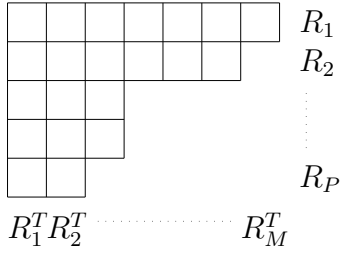}
\end{center}
\caption{A Young tableau with $P\leq N$ rows and $M$ columns labeling a representation of $U(N)$. $R_i$ is the number of boxes in the $i$-th row and satisfies $R_i\geq R_{i+1}$. $R^T$ is the tableau conjugate to $R$, obtained by exchanging rows with columns.}
\label{young-param-1}
\end{figure}

We show that the information about the representation $R$ of the
Wilson loop operator -- given by a Young tableau (see Figure
\ref{young-param-1}) -- is encoded in the eigenvalues\footnote{
Geometrically, these eigenvalues correspond to the positions of the
branes up to Hamiltonian deformations (see Appendix
\ref{target-theory}).} of the holonomy matrix obtained by
integrating the gauge connection on the branes wrapping $L$ --
denoted by ${\cal A}$ -- around the non-contractible
$\beta$-cycle\footnote{ $L$ has the topology of a solid torus with a
boundary at infinity given  by a $T^2$ with canonical $\alpha$ and
$\beta$ one cycles. The $\beta$ cycle is non-contractible in $L$
while $\alpha$ may or may not be contractible.}  of the Lagrangian
submanifold $L$.

We show that a Wilson loop labeled by a knot $\alpha$ and a
representation $R$ given by Figure \ref{young-param-1} is described
either by  a configuration of $M$ D-branes
$(L_{x_1},L_{x_{2}},\ldots,L_{x_M})$ or by a configuration of
$P$ anti-branes $({\bar L}_{y_1},{\bar L}_{y_{2}},\ldots,{\bar
L}_{y_P})$, where $L_{x_i}/{\bar L}_{y_i}$ denotes  a
D-brane/anti-brane with world-volume   $L$ and with a non-trivial
holonomy, shifted by the integral of the \Kahler form $\Jcal$,
\ba
&&x_i\equiv \oint_{\beta=\p D} {\cal A}_i+\int_D \Jcal=g_s\left(R^T_i-i+M+\half\right),~
i=1,\dots,M,\\
&& y_i\equiv \oint_{\beta=\p D} {\cal A}_i+\int_D
\Jcal=g_s\left(R_i-i+P+\half\right),~ i=1,\dots,P, \ea
respectively\footnote{$\Jcal$ is integrated over a disk $D$.  $D$
is a relative cycle in the resolved conifold and has $\beta\subset
L$ as its boundary.}.

For the case of the simplest knot -- the unknot -- we explicitly
check this identification. We show that this Wilson loop operator in
a representation given by Figure \ref{young-param-1} corresponds to
the configuration of $M$ D-branes $(L_{x_1},L_{x_2},\ldots,L_{x_M})$
 \begin{figure}[h]
 \begin{center}
\psfrag{x1}{$x_1$}
\psfrag{x2}{$x_2$}
\psfrag{xM}{$x_M$}
\includegraphics[width=60mm]{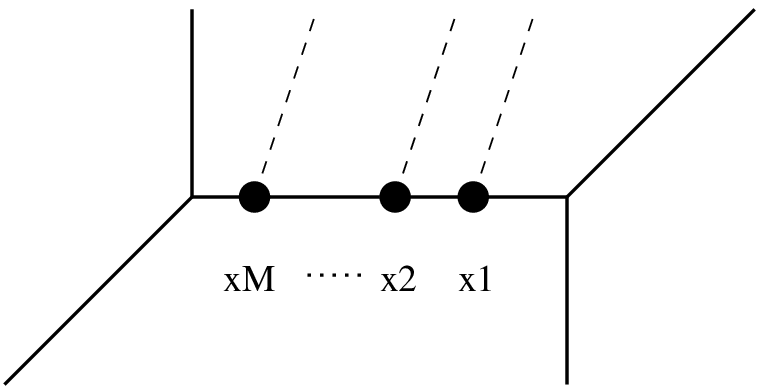}
\end{center}
\caption{Brane configuration in resolved conifold describing Wilson loop in a representation given by Figure \ref{young-param-1}.}
\label{brane-inner_2}
\end{figure}
\newline
\noindent where $L_{x_i}$ denotes a D-brane wrapping a Lagrangian
submanifold $L$  sitting at the position $x_i=g_s(R^T_i-i+M+1/2)$ on
the inner edge of the toric  diagram of the resolved conifold. In
particular, a single D-brane on the inner edge corresponds to a
Wilson loop in the antisymmetric representation. Since the effective
size of the inner edge of the resolved conifold is
$t_{eff}=g_s(N+1)$, we recover (from $x_1\leq t_{eff}$) the group
theory bound $R^T_1\leq N$   for the number of boxes in a column
from the compactness of the $\CP^1$ of the resolved conifold.

The same Wilson loop operator can also be described by a
configuration of $P$ anti-branes
$(\bar{L}_{y_1},\bar{L}_{y_2},\ldots,\bar{L}_{y_P})$
\begin{figure}[htbp]
 \begin{center}
\psfrag{y1}{$y_1$}
\psfrag{y2}{$y_2$}
\psfrag{yP}{$y_P$}
\includegraphics[width=40mm]{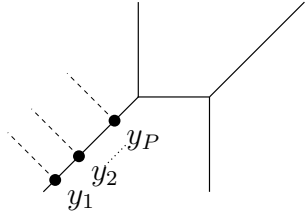}
\end{center}
\caption{Brane configuration in resolved conifold describing Wilson loop in a representation given by Figure 1.}
\label{brane-outer_2}
\end{figure}
\newline
\noindent where $\bar{L}_{y_i}$ denotes an anti-brane wrapping a
Lagrangian submanifold $L$  sitting at the position
$y_i=g_s(R_i-i+P+1/2)$ on an outer edge of the toric diagram. In
particular, a single D-brane on the outer edge corresponds to a
Wilson loop in the symmetric representation. Since the outer edge is
non-compact, $R_1$ can be an arbitrarily large integer, as expected
from group theory. The bound on the number of anti-branes $P\leq N$
can also be understood geometrically. Once we put $P$ anti-branes in
the resolved conifold, the effective size of the space is
$t_{eff}=g_s(N-P)$, thus recovering the group theory bound $P\leq
N$.

These results are obtained by computing the A-model topological string partition function in the presence of these branes and showing that it agrees with the Chern-Simon results of Witten \cite{Witten:1992fb}.
In the crystal melting description \cite{Okounkov:2003sp,Okuda:2004mb}  of these amplitudes, the Young tableau of the dual Wilson loop has a nice geometrical realization in the crystal. The corners produced in the crystal by the insertion of branes \cite{Saulina:2004da,Okuda:2004mb} has exactly the shape of the corresponding Young tableau!

We also find a purely geometric description of the unknot Wilson loop operators in terms of  Calabi-Yau geometries that are asymptotic to the resolved conifold. We  show that the backreaction of the brane configurations  in
Figures 2 and 3 can be taken into account exactly,   giving rise to a Calabi-Yau geometry without D-branes!  The two different D-brane configurations in Figures 2 and 3 yield the same Calabi-Yau geometry, and the  information about the representation $R$ of the Wilson loop is now encoded in the topology of the corresponding Calabi-Yau. Physically, the appearance of this rich class of Calabi-Yau geometries can be understood as arising from a new class of geometric transitions, where non-compact Lagrangian D-branes/anti-branes are replaced by non-trivial topologies supporting the ``flux" produced by the branes.
The computation of the A-model topological string amplitude on these ``bubbling" Calabi-Yau geometries, obtained by letting the branes undergo a geometric transition, exactly reproduces the result for the expectation value of the corresponding Wilson loop obtained by Witten \cite{Witten:1992fb}. Therefore, these bubbling Calabi-Yau geometries
 provide  a novel representation of knot invariants of Chern-Simons theory.

The topology of the  Calabi-Yau geometry corresponding to a Wilson loop  is best understood by giving the following parametrization of a Young tableau\footnote{Informally, $l_{odd}$ is the number of rows
in the tableau with the same number of boxes while $l_{even}$ is the number of columns in the tableau with the same number of boxes.}
\begin{figure}[htbb]
\centering
\begin{tabular}{cc}
\psfrag{N}{$N$}
\psfrag{l1}{$l_1$}
\psfrag{l2}{$l_2$}
\psfrag{l2m-1}{$l_{2m-1}$}
\psfrag{l2m}{$l_{2m}$}
\psfrag{l2m+1}{$l_{2m+1}$}
\includegraphics[width=70mm]{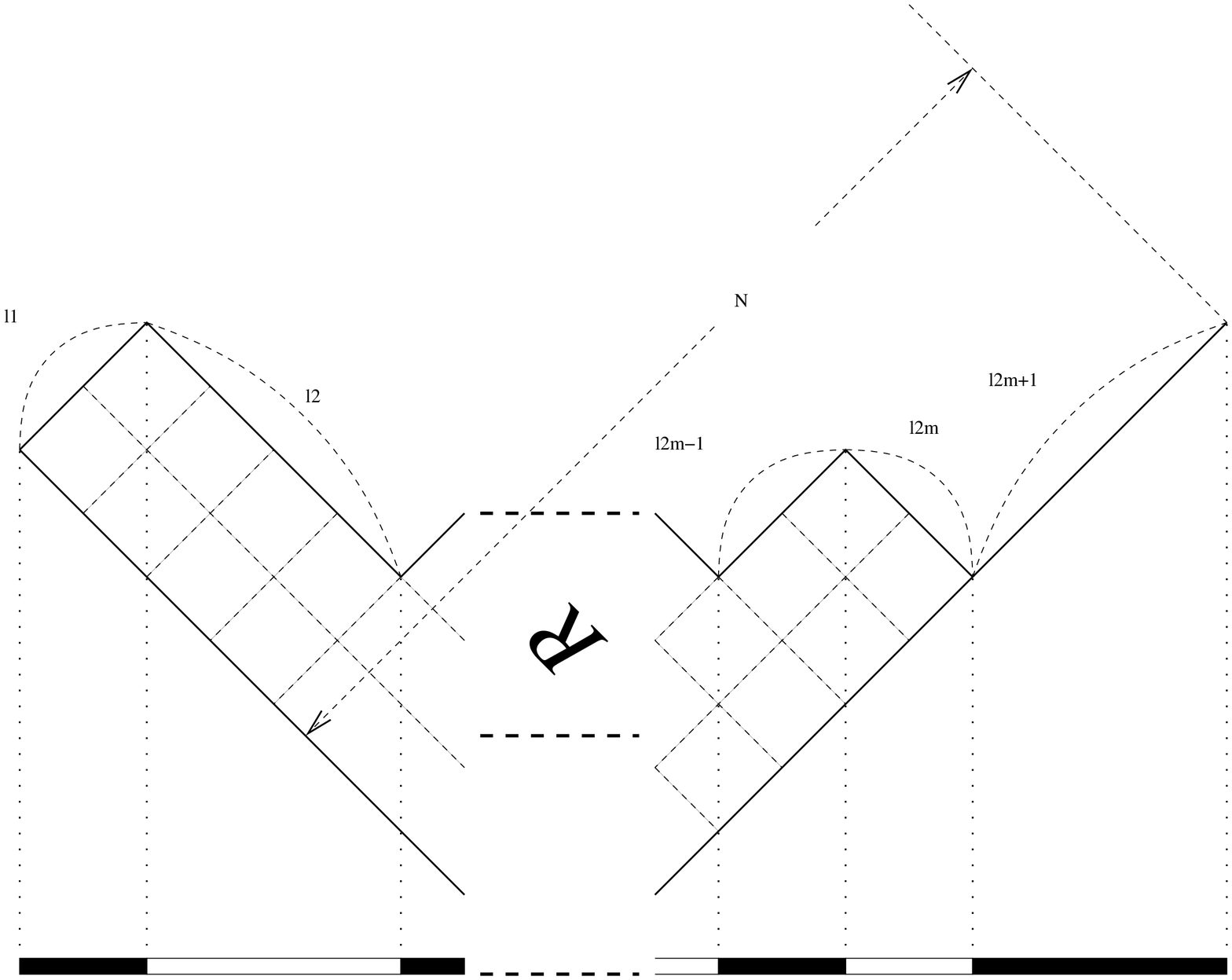}&
\psfrag{t1}{$t_1$}
\psfrag{t2}{$t_2$}
\psfrag{t3}{$t_3$}
\psfrag{t2m}{$t_{2m}$}
\psfrag{t2m+1}{$t_{2m+1}$}
\includegraphics[width=70mm]{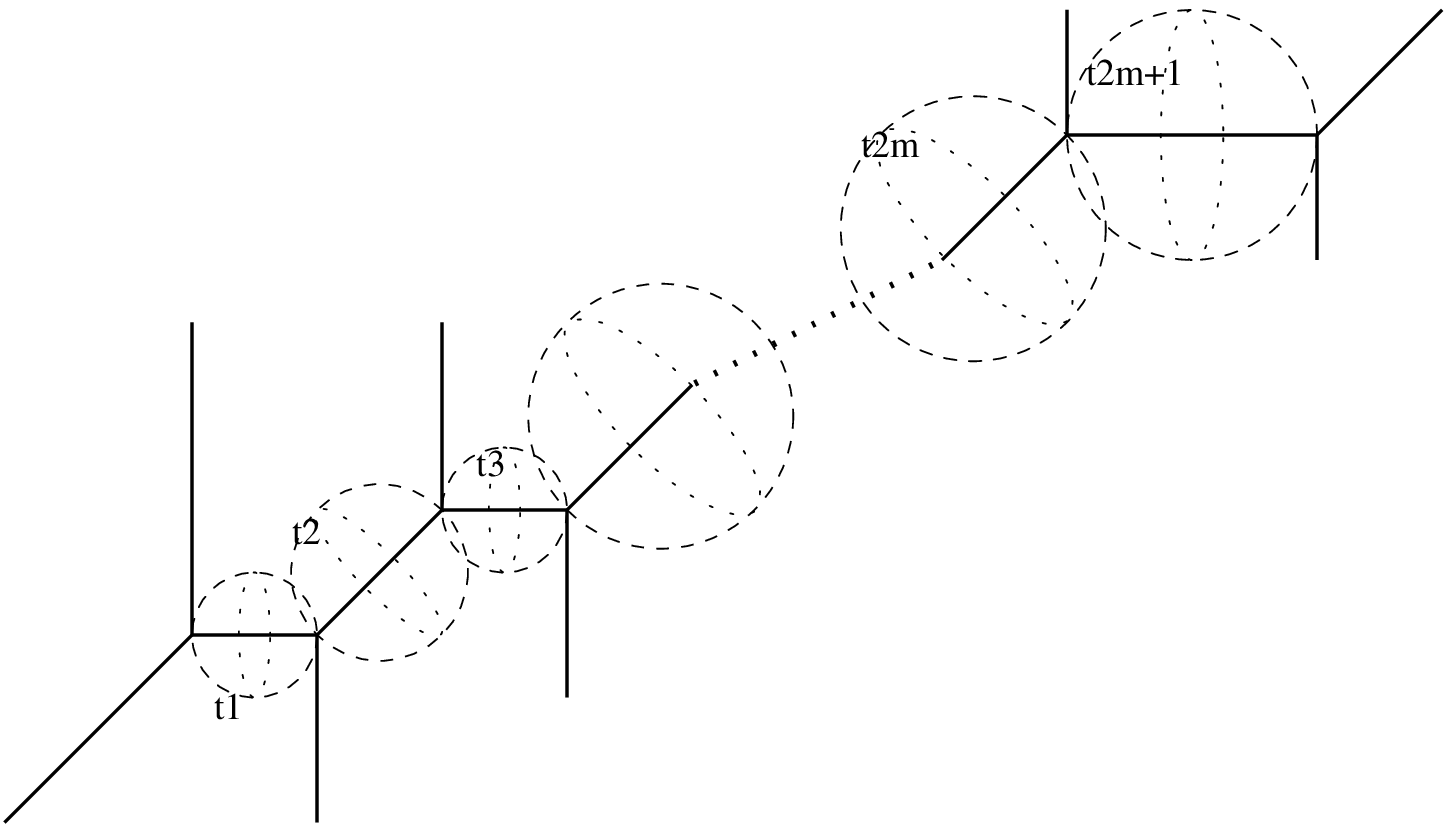}\\
(a)&(b)
\end{tabular}
\caption{
The correspondence of the Wilson loop in representation $R$ and the bubbling
geometry.
In (a) the Young tableau $R$, shown rotated, is specified by the lengths $l_i$ of all the
edges. $l_{2m+1}$ is $N$ minus the number of rows.
Equivalently, $l_i$ are the lengths of black and white regions in
the Maya diagram.
The Wilson loop in  representation $R$
(a) is equivalent to the toric Calabi-Yau manifold given by the web
diagram (b).
There are a total of $2m+1$ bubbles of $\CP^1$ in the geometry.
The sizes of the $\CP^1$'s are given by $t_i=g_sl_i,~i=1,...,2m+1$.
}
\label{bubbling-geometry_2}
\end{figure}

The Calabi-Yau geometry dual to a Wilson loop labeled by a Young tableau given  in Figure \ref{bubbling-geometry_2}(a) is
given by the toric web diagram shown in Figure \ref{bubbling-geometry_2}(b).
The corresponding Calabi-Yau geometry  has $2m+1$ nontrivial $\CP^1$'s. The associated complexified \Kahler moduli are given by $t_i=g_sl_i$ for $i=1,\ldots, 2m+1$, where $l_i$ are the integers in Figure \ref{bubbling-geometry_2}(a) parametrizing the Young tableau.
Similarity between bubbling AdS geometries and toric geometries was noted in \cite{Lin:2004nb}.

Although the explicit quantitative checks we make are all based on
the unknot, our physical arguments apply to arbitrary knots and
links. Indeed it is possible to write the closed Gopakumar-Vafa (GV)
invariants \cite{Gopakumar:1998ki} associated to the bubbling Calabi-Yau
corresponding to an arbitrary knot in terms of the
 open GV invariants \cite{Ooguri:1999bv,Labastida:2000yw} associated to  the
Lagrangian branes giving rise to the bubbling Calabi-Yau after  geometric transition. We plan to
report on this relation and on more general geometric transitions in
the forthcoming paper \cite{geom-trans}.

The extent of the analogy between AdS/CFT and the GV duality
is striking.
We hope that the very explicit realization of the holographic
correspondence found in this paper -- including all quantum corrections --  provides us with hints for deeper understanding of
the AdS/CFT correspondence and holography in further detail.

The plan of the rest of the paper is as follows. In section $2$ we
integrate out the degrees of freedom introduced by a configuration
of non-compact branes intersecting ${\rm S}^3$ along a knot $\alpha$
and show that the net effect is to insert a Wilson loop operator for
$U(N)$ Chern-Simons theory on ${\rm S}^3$. In section $3$ we
identify the brane configurations in the resolved conifold dual to a
Wilson loop operator. The identification is explicitly checked for
the case of the unknot. In section $4$ we show that the Wilson loop
operators can be related to the closed string partition function of
bubbling Calabi-Yau's, which are interpreted as arising via a
geometric transition of the brane configuration in section $3$.
Appendix \ref{F1} revisits the description of Wilson loops in terms
of string world-sheets. Appendix \ref{crystal-section} reviews the
melting crystal models found in \cite{Okuda:2004mb}.  The explicit
computations for the unknot, which support our proposal in this
paper, are performed using the crystal models. Appendix
\ref{target-theory} explains relevant details of the target space
theory of open topological strings.

\section{Wilson loops as  branes in the deformed conifold}

In this section we find a brane configuration in $T^*{\rm S}^3$   whose effective description is given by a Wilson loop operator of Chern-Simons theory on ${\rm S}^3$. The basic idea is to
 add extra non-compact branes  beyond  the $N$ D-branes wrapping the ${\rm S}^3$ that support the $U(N)$ Chern-Simons theory participating in the large $N$ duality. We show that integrating out the degrees of freedom introduced by the extra branes has the effect of inserting into the
$U(N)$ Chern-Simons path integral on ${\rm S}^3$ a Wilson loop operator in a particular representation $R$ of $U(N)$ determined by data of  the brane configuration. Having identified the brane configuration in $T^*{\rm S}^3$ corresponding to a Wilson loop, in the next section we find the holographic description of Wilson loops in terms of branes in the resolved conifold.

A Wilson loop operator of  Chern-Simon theory on ${\rm S}^3$ is labeled by a representation $R$  of  $U(N)$ and by an embedded oriented circle $\alpha$  -- also known as a knot --  in ${\rm S}^3$.
 It is given by
\ba
W_R(\alpha)=\hbox{Tr}_R P \exp\oint_\alpha A,
\label{wilsonopp}
\ea
where $P$ is the path ordering operator.

A microscopic  description of a Wilson loop operator can be given  by adding new degrees of freedom localized on the knot on which the operator is supported.  In the string theory realization, the  new localized degrees of freedom arise microscopically by quantizing open strings between the D-branes wrapping  ${\rm S}^3$ and new  branes intersecting the ${\rm S}^3$ along the knot $\alpha$.

As shown by \cite{Ooguri:1999bv}, a Lagrangian submanifold  $L$ in   $T^*{\rm S}^3$
can be found such that it intersects   ${\rm S}^3$ along an arbitrary knot $\alpha$. Given a knot $\alpha$ in ${\rm S}^3$ parametrized  by $q^i(s)$, where $q^i$ are coordinates on ${\rm S}^3$, then $L$ is determined  by the following non-compact Lagrangian submanifold\footnote{We recall that the \Kahler form of $T^*{\rm S}^3$ is given by $\omega=\sum_{i=1}^3dp_i\wedge dq^i$.}
\ba
L=\{(q^i,p_i) | \sum_{i}^3 p_i \f{dq^i}{ds}=0\},
\label{lagrangian}
\ea
where $p_i$ are coordinates in the fiber of $T^*{\rm S}^3$. The topology of $L$ is ${\rm R}^2\times {\rm S}^1$ or more precisely that of a solid torus. The brane is non-compact and  has a boundary at asymptotic infinity. The boundary is   a $T^2$ with canonical $\alpha$ and $\beta$ one cycles.
In $L$,
 the $\alpha$-cycle of the $T^2$ is non-contractible  while the $\beta$-cycle of the $T^2$ is   contractible\footnote{As we will see in the next section,  the roles of $\alpha$ and $\beta$ are reversed in the corresponding Lagrangian submanifold $L$ in the resolved conifold;
 $\beta$ being non-contractible in the corresponding $L$.}.

Therefore, we want to consider the gauge theory living on the brane configuration  in $T^*{\rm S}^3$ given by $N$ D-branes wrapping  ${\rm S}^3$ intersecting $M$ branes wrapping $L$. The action is given by \cite{Ooguri:1999bv}
\ba
S=S_{CS}(A)+S_{CS}({\cal A})+\oint_\alpha   \chi^\dagger (d+A+{\cal A})\chi,
\label{path}
\ea
where
\ba
S_{CS}(A)=\f{1}{g_s}\int_{X} \hbox{Tr}\left(A\wedge dA+\f{2}{3} A\wedge A \wedge A \right),\cr
\ea
and $A$, ${\cal A}$ are the connections on $X={\rm S}^3$ and $X=L$ respectively. $\chi$ are bifundamental fields localized along the knot $\alpha$ arising from the quantization of open strings with one end on ${\rm S}^3$ and the other one on $L$. Depending on whether we wrap a D-brane or an anti-brane\footnote{In  topological string theory an anti-brane is exactly the same \cite{Vafa:2001qf,Okuda:2006fb} as a ghost brane, obtained by reversing  the sign of the D-brane boundary state. That's why the statistics of the open string fields is reversed when open strings are stretched between a D-brane and an anti-brane as compared to when they are stretched between two D-branes. } on $L$, $\chi$ is either a fermionic or bosonic field.

\subsection{Wilson loops as D-branes}

If D-branes wrap $L$ then $\chi$ are fermions. Integrating out $\chi$ is straightforward,
it inserts the so called Ooguri-Vafa operator \cite{Ooguri:1999bv}\footnote{
The path integral over $\chi$  reduces to that of free fermions.
This formula can then be derived by
noting that
$\prod_{i=1}^N\prod_{I=1}^M(1+ x_iy_I)=\sum_{Q} \Tr_{Q^T}X\Tr_QY$, where $X={\rm diag}(x_i),~Y={\rm diag}(y_I)$.}
\ba
\sum_{Q}\hbox{Tr}_{Q^T} P \exp\oint_\alpha A\cdot \hbox{Tr}_{Q} P \exp\oint_\alpha {\cal A},
\ea
where $Q^T$ is the Young tableau conjugate to $Q$.  We now want to integrate ${\cal A}$ out\footnote{In this analysis we omit the path integral over $A$, which is to be done at the end. In \cite{Ooguri:1999bv} the path integral over $A$ was performed while the integral over ${\cal A}$ wasn't.}
\ba
e^{iS_{CS}(A)}\sum_{Q}\hbox{Tr}_{Q^T} P \exp\oint_\alpha A\cdot \int[D{\cal A}] e^{iS_{CS}({\cal A})}  \hbox{Tr}_{Q} P \exp\oint_\alpha {\cal A}
\label{pathint}
\ea

In order to proceed we now briefly recall some well known facts about Chern-Simons theory. In the present case, Chern-Simons is defined on a solid torus with a boundary at infinity given by a $T^2$.
The path integral is a wave function $\langle \psi |$ for Chern-Simons theory on a $T^2$ and
depends on the boundary condition imposed at infinity.

The path integral in (\ref{pathint}) has the insertion of $\hbox{Tr}_{Q} P \exp\oint_\alpha {\cal A}$, that is a Wilson loop along the non-contractible cycle $\alpha$ of $T^2$. This creates a state of Chern-Simons theory on $T^2$ labeled by $|Q\rangle$. Therefore (\ref{pathint}) yields
 \ba
e^{iS_{CS}(A)}\sum_{Q}\hbox{Tr}_{Q^T} P \exp\oint_\alpha A \cdot
\langle\psi|Q\rangle. \label{pathint-brane}
\ea

The holonomies around the $\alpha$ and $\beta$ cycles of $T^2$ are
canonically conjugate \cite{Elitzur:1989nr}. The path integral is
defined by specifying the holonomy at infinity either around the
$\alpha$ or $\beta$ cycle. Roughly speaking, the holonomy around
$\alpha$ plays the role of the position operator while the holonomy
around the $\beta$ cycle plays the role of momentum. In the context
of Chern-Simons theory,   momentum can be identified with the
highest weight vector  of a representation $R$, shifted by the Weyl vector.

Since our aim is to find  the Wilson loop in a particular
representation $R$ (\ref{wilsonopp}), we choose the boundary condition $\langle \psi |= \langle R^T|$. This means that we choose a boundary condition at infinity with non-trivial holonomy around the contractible cycle $\beta$ in $L$. Since the
states labeled by representations form an orthonormal basis, this
picks out the $Q=R^T$ term in (\ref{pathint-brane}). In other words,
imposing the boundary condition $\langle \psi|=\langle R^T|$ is
equivalent to focusing on the $Q=R^T$ term in (\ref{pathint-brane}).

We now give the physical interpretation of this boundary condition. In a nutshell, the holonomy around the contractible cycle $\beta$ measures fundamental string charge.

\medskip
\medskip
\medskip
\noindent {\it Open String Endpoints as Anyons }

\medskip
\medskip
\medskip
To understand the physical meaning of the boundary condition, let us
consider the effect of the Wilson loop $\Tr_{R}\exp\oint \Acal$ on
the non-compact branes.

Consider first the case when $M=1$, i.e. for a single brane on $L$.
The   action of the brane in the presence of the Wilson loop (swept by an open string end point) is given by
\ba
S=S_{CS}(\Acal)+k\oint_\alpha \Acal.
\ea
Therefore, the corresponding equations of motion are given by
\ba
F_{z\bar z}=g_sk \delta^2(z),\label{vortex}
\ea
where $z$ is the complex coordinate parameterizing the ${\rm R}^2$ in $L$.
Physically, the endpoint of an open string behaves like an
anyon when  viewed from the  brane world-volume.

By integrating (\ref{vortex}) we conclude that having $k$ fundamental strings ending on $L$ introduces a non-trivial holonomy around the contractible cycle $\beta$ of the $T^2$  in $L$. The holonomy is given by
\ba
\oint_\beta {\cal A} =g_s k.
\ea

In the general case of $M$ arbitrary, the treatment of the
non-Abelian equations of motion is more subtle, as noted originally
by Witten \cite{Witten:1988hf}. Fortunately, we can borrow results
from \cite{Elitzur:1989nr}, where the insertion of the Wilson loop
in the representation $R^T$ was found to induce the holonomy \ba
\oint_\beta \Acal_i=g_s\left(R^T_i-i+\half
M+\half\right),~i=1,2,...,M.
\label{holo} \ea
Therefore, the boundary condition $\langle \psi|=\langle R^T|$ is tantamount to introducing the holonomy given by (\ref{holo}) at infinity.
We denote by $(L_{  R^T_1},\ldots,L_{R^T_M})$ the
configuration of branes with this holonomy (\ref{holo}).

When we go through the transition to the resolved conifold, we will
see that there is a shift by $\half g_s M$ in the effective
holonomy, coming from the backreaction of the geometry on the
branes. In Appendix C we show that the gauge invariant quantity is
given by the holonomy shifted by the integral of the \Kahler form
over a disk ending on $\beta$, thus giving rise to the shift
by\footnote{The shift of the \Kahler modulus due to the insertion of
branes is well known
\cite{Aganagic:2003db,Saulina:2004da,Okuda:2004mb} and it is given
by $g_sM$. Therefore, when the \Kahler modulus is integrated over a
disk bounding $\beta$ -- as opposed to over $\CP^1$ -- we get a
shift by $\half g_s M$.
\label{footnote-shift}} $\half g_s M$ in the resolved conifold.

Therefore, the D-brane configuration $(L_{  R^T_1},\ldots,L_{
R^T_M})$ we have considered in $T^*{\rm S}^3$ inserts a Wilson loop
operator along the knot $\alpha$
 \ba
e^{iS_{CS}(A)} \cdot \hbox{Tr}_{R} P \exp\oint_\alpha A.
\ea
Summarizing, we have derived the identification \ba (L_{
R^T_1},\ldots,L_{  R^T_M})\leftrightarrow \hbox{Tr}_{R} P
\exp\oint_\alpha A. \ea

\subsection{Wilson loops as anti-branes}

In topological string theory an anti--brane  has the interpretation as a ghost brane
\cite{Vafa:2001qf,Okuda:2006fb}, whose boundary state is minus that of a brane.
This sign changes the statistics of the open string fields between branes and anti-branes.
This is why the $\chi$ fields in the brane configuration
we have described in $T^*{\rm S}^3$ are bosonic when anti-branes wrap the Lagrangian submanifold $L$.

When $\chi$ are quantized as bosonic fields, integrating them out yields\footnote{
Now the path integral over $\chi$ reduces to that of free bosons.  The final answer is obtained by noting that
$\prod_{i=1}^N\prod_{I=1}^M\oo{1- x_iy_I}=\sum_{Q} \Tr_{Q}X\Tr_QY)$.}
\ba
\sum_{Q}\hbox{Tr}_{Q} P \exp\oint_\alpha A\cdot \hbox{Tr}_{Q} P \exp\oint_\alpha {\cal A}.
\ea
We can now easily calculate the path integral over ${\cal A}$
following our previous  discussion for D-branes. If we denote by $({\bar L}_{  R_1},\ldots,{\bar L}_{  R_P})$
a configuration of $P$ anti-branes with
holonomy\footnote{As before,
there will be a shift by $\half g_s P$ in the resolved conifold.}

\ba
\oint_\beta \Acal_i=g_s\left(R_i-i+\half P+\half\right),~i=1,2,...,P,
\label{taku}
\ea
corresponding to the boundary condition $\langle \psi| =\langle R|$,
we are left with the insertion of a Wilson loop operator
along the knot $\alpha$ given by
 \ba
e^{iS_{CS}(A)} \cdot \hbox{Tr}_{R } P \exp\oint_\alpha A.
\ea
Therefore we arrive at the identification
\ba
({\bar L}_{  R_1},\ldots,{\bar L}_{  R_P})\leftrightarrow \hbox{Tr}_{R} P \exp\oint_\alpha A.
\ea

We now study the bulk description of Wilson loops in Chern-Simons theory in terms of branes in the resolved conifold geometry.

\section{Wilson loops as  branes in the resolved conifold} \label{branes-res}

In the previous section we have shown that a Wilson loop operator on
any knot $\alpha$ and for any representation $R$ can be obtained by
integrating out the physics of a D-brane configuration  $(L_{
R^T_1},\ldots,L_{  R^T_M})$ or anti-brane configuration  $(\bar{L}_{
R_1},\ldots,\bar{L}_{  R_P})$.

To obtain the resolved conifold description of a Wilson loop we follow the brane configuration through the conifold singularity.
It is possible to construct a Lagrangian submanifold $L$ explicitly for every knot
$\alpha$ in ${\rm S}^3$ \cite{Taubes:2001wk,Koshkin}. Physically, this Lagrangian submanifold in the resolved conifold is the dual description of the Lagrangian submanifold (\ref{lagrangian}) in $T^*{\rm S}^3$  described in the previous section. Topologically $L$ is also ${\rm R}^2\times {\rm S}^1$ and has
an asymptotic boundary given by a $T^2$.

The Lagrangian submanifolds constructed by Taubes
have the property that they end on a knot $\alpha$ in the ${\rm S}^3$ at infinity. From the point of view of holography, this is precisely   as expected. The dual $U(N)$ Chern-Simons theory lives on the ${\rm S}^3$ at asymptotic infinity in the resolved conifold. Given that we are looking for the resolved conifold description of Wilson loops, it is expected that the bulk description is given by a bulk object which ends on a knot, thus introducing the appropriate source  for a Wilson loop operator.

A crucial role in the derivation in the previous section is played by the holonomies
around the contractible cycle of the $T^2$ in $L$
\ba
\oint_\beta {\cal A}_i.
\label{holoa}
\ea
This data has a nice geometrical interpretation in the resolved conifold. As one follows the branes through the conifold singularity, the contractible cycle $\beta$ of the Lagrangian $L$ in $T^*{\rm S}^3$ grows to become a non-contractible cycle $\beta$ of the $T^2$ on the corresponding Lagrangian $L$  of the resolved conifold.

As explained in more detail in  Appendix C, the holonomy of the
complex
gauge field ${\cal A}$
has the interpretation as the modulus\footnote{In the toric diagram,
the modulus is the position of the brane.} of the brane. More
precisely, in addition to the holonomy of the gauge field, the gauge invariant modulus
has a contribution from the \Kahler form $\Jcal$ integrated over a
disk ending on $\beta$. This contributes  $\half g_s M$ to the modulus
since $\int_D \Jcal =\half g_s M$ (see footnote \ref{footnote-shift}).

Therefore, the Wilson loop operator described by the brane
configuration $(L_{R^T_1},\ldots,L_{R^T_M})$ in the deformed
conifold has a bulk interpretation in terms of $M$ Lagrangian D-branes
$(L_{x_1},\ldots,L_{x_M})$ in the resolved conifold. Moreover, the
modulus $x_i$ of the $i$-th brane in the resolved conifold is determined by the holonomy data
(\ref{holo}) \ba x_i=\oint_{\p D=\beta} \Acal+\int_D \Jcal=g_s\left(
R^T_i-i+{M } +\half\right)\qquad i=1,\ldots,M\ea

Similarly, the brane configuration $({\bar L}_{  R_1},\ldots,{\bar L}_{  R_P})$ in the deformed
conifold has a bulk interpretation in terms of $P$ Lagrangian anti-branes
$(\bar L_{y_1},\ldots,\bar L_{y_P})$ in the resolved conifold. Moreover, the
modulus $y_i$ of the $i$-th brane in the resolved conifold is determined by the holonomy data
(\ref{taku}) \ba y_i=\oint_{\p D=\beta} \Acal+\int_D \Jcal=g_s\left(
R_i-i+{P } +\half\right)\qquad i=1,\ldots,P \ea

We now proceed to verify this identification for the case when
the Wilson loop operator is defined on the simplest knot, the unknot.
When the deformed conifold geometry is defined by
\ba
z_1 z_4-z_2 z_3=\mu, \label{def-con}
\ea
the unknot is parameterized by $(z_1,z_2,z_3,z_4)=(0,\sqrt \mu
e^{i\theta},-\sqrt\mu e^{-i\theta},0)$ with $0\leq \theta\leq 2\pi$.

For the case of the unknot, the Wilson loop operator $W_{R}$ in (\ref{wilsonopp}) is described by the configuration of D-branes in
Figure
\ref{brane-inner_2}\footnote{
Let us parameterize the resolved conifold by
$|\lambda_1|^2+\lambda_2|^2-|\zeta_1|^2-|\zeta_2|^2={\rm Re}\hspace{.5mm}t$ up to $U(1)$
equivalence with charges $(1,1,-1-1)$.
In Figures \ref{brane-inner_2} and \ref{brane-outer_2},
the left, right, top, and bottom regions correspond to $|\lambda_1|^2=0,
|\lambda_2|^2=0, |\zeta_1|^2=0,$ and $|\zeta_2|^2=0$, respectively.
$L$ is given by $
|\lambda_1|^2-{\rm Re}\hspace{.5mm}x=|\lambda_2|^2-{\rm Re}\hspace{.5mm}(t-x)=|\zeta_1|^2=|\zeta_2|^2,
~\arg(\zeta_1\zeta_2\lambda_1\lambda_2)=\pi$.
When the conifold is singular the $z$ and $(\lambda,\zeta)$ coordinates are related as
$
\left(
  \begin{array}{cc}
    z_1 & z_2 \\
    z_3 & z_4 \\
  \end{array}
\right)=\left(\begin{array}{c}
\zeta_1\\
\zeta_2
\end{array}\right)
\left(\begin{array}{cc}
-\lambda_2&\lambda_1
\end{array}\right).
$
\label{res-con-convention}
}.
The corresponding Lagrangian submanifolds end on the inner edge, thus accounting for the group theory
bound $R^T_1\leq N$.
This is consistent with the requirement that $\beta$ should be
non-contractible.

It was shown in \cite{Okuda:2004mb}, as reviewed in Appendix
\ref{crystal-section}, that the Wilson loop vev for the unknot can
be rewritten as \ba \langle W_R\rangle=
M(q)e^{-\sum_{n=1}^\infty\frac{e^{-nt}}{n[n]^2}}
\left(\prod_{i<j}(1-e^{-(x_i-x_j)})\right) \prod_{i=1}^M \exp
\sum_{n=1}^\infty \frac{e^{-nx_i}+e^{-n({t}-x_i)}}{n[n]} ,
\label{brane-amplitude} \ea up to unimportant factors we suppress
here. Here $M(q)=\prod_{j=1}^\infty (1-q^j)^{-j}$ is the MacMahon function,
$t=g_s(N+M)$ and $x_i=g_s(R^T_i-i+ M+\half),
~~i=1,...,M.$ This is exactly the A-model amplitude for the brane
configuration in Figure 2\footnote{The factor $\prod_{i<j}(1-e^{-(x_i-x_j)})$ is the
contribution from annulus diagrams between branes.
This is essentially the Vandermonde determinant, i.e. the Weyl denominator.
It demonstrates that
the branes are fermions, as shown first in the mirror B-model \cite{Aganagic:2003db}.
The rest of (\ref{brane-amplitude}),
up to $M(q)$, can be computed by the topological vertex technology \cite{Aganagic:2003db}.},
thus confirming our identification \ba
\langle W_R\rangle =\langle (L_{x_1},\ldots,L_{x_M})\rangle . \ea

We also show in Appendix \ref{crystal-section} that the Wilson loop
vev can also  be written as \ba \langle W_R\rangle= M(q)
e^{-\sum_{n=1}^\infty\frac{e^{-n{t}}}{n[n]^2}} \left(\prod_{
i<j}(1-e^{-(y_i-y_j)})\right) \prod_{i=1}^P \exp
\sum_{n=1}^\infty \frac{e^{-ny_i}-e^{-n({t}+y_i)}}{n[n]} ,
\label{anti-brane-amplitude} \ea with
$y_i=g_s(R_i-i+P+\half),~i=1,...,P,~{t}=g_s(N-P)$. This is exactly the A-model
amplitude for the brane configuration in Figure 3, thus confirming
our identification\footnote{
The same amplitude would arise if the branes ended on any of the four outer edges.
One can convince oneself that they end on the lower-left edge by the following argument.
In the convention of footnote \ref{res-con-convention}, one can show that $\alpha$ is the
 trivial cycle in $L$
if $L$ ends on the lower-left or upper-right edge.
$\beta$ is trivial if $L$ ends on the lower-right or upper-left edge.
Since we have non-trivial holonomy of a flat connection along $\beta$,
$\beta$ has to be non-trivial, and thus $L$ has to end on either the lower-left or upper-right edge.
By carefully keeping track of the orientation of the cycle $\beta$, one can show that
increasing the holonomy is gauge equivalent to moving the branes toward the lower-left.
Since one can increase the holonomy without bound, the branes must end on the lower-left edge.
Such $L$ is
given by the equations $ |\lambda_1|^2=|\lambda_2|^2-{\rm
Re}\hspace{.5mm}(t+y)=|\zeta_1|^2-{\rm
Re}\hspace{.5mm}y=|\zeta_2|^2,
~\arg(\zeta_1\zeta_2\lambda_1\lambda_2)=\pi$. } \ba \langle
W_R\rangle =\langle({\bar L}_{y_1},\ldots,{\bar L}_{y_P})\rangle.
\ea

In summary we have  identified the bulk description of Wilson loop operators in Chern-Simons theory in terms of D-branes and anti-branes in the resolved conifold. Moreover, this identification has been explicitly verified for the case of the unknot.

\section{Wilson loops as bubbling Calabi-Yau's}
\label{section-bubbling}

A geometric transition is a phenomenon in which a stack of D-branes is
replaced by a new geometry with ``flux''.
More precisely, when the number of branes in the stack is large,
the system is better described by a certain geometry where the appropriate
fields that encode the charges of the branes are turned on.
In physical string theory these fields are RR fluxes originally sourced by the D-branes.
In topological string theory, the role of the fluxes is played by bulk gauge fields, namely the \Kahler form and the B-field.
The change in the geometry is such that the non-trivial cycle, which originally surrounds\footnote{
Let $W$ be the world-volume of the brane, and $M$ a homologically
trivial cycle.
We say that $M$ surrounds $W$ if there is a chain $N$ such that $\p
N=M$ and if the  intersection number of $N$ and $W$ is non-zero.
}
the branes
and is homologically trivial, becomes a non-trivial cycle that supports the ``flux''.
There are by now many examples of this phenomenon.
Here we will find a new class of geometric transitions in topological string theory.

It has been argued in \cite{Yamaguchi:2006te,Lunin:2006xr} that the D-branes
realizing the half BPS straight Wilson lines
in $\Ncal=4$ SYM can undergo  transition
to certain {\it bubbling} geometries that asymptote
to $AdS_5\times {\rm S}^5$.
To obtain these geometries
one makes an ansatz for supergravity fields based on the knowledge of the symmetry
of the branes describing the Wilson loops.
One then imposes the BPS condition and finds that the supergravity solution is
determined by simple data, namely a black-and-white pattern on a line as in Figure \ref{bubbling-geometry_2}(a).
It is expected that the data corresponds to
the representation of the Wilson loop \cite{Yamaguchi:2006te}.

We found in section \ref{branes-res} that Wilson loop
operators in Chern-Simons theory can be realized by a configuration
of branes or anti-branes in the resolved conifold.
When the number of D-branes in a stack is large we expect that the system has
a better description in terms of pure geometry.
The D-branes wrap a Lagrangian submanifold $L$ of topology ${\rm R}^2\times {\rm S}^1$.
The neighborhood of the submanifold, modeled by the normal bundle, is locally $\R^5\times {\rm S}^1$.
A contractible ${\rm S}^2$ in the transverse $\R^3$ surrounds the branes.
The geometric transition of the D-branes makes the
${\rm S}^2$ non-trivial while making the ${\rm S}^1$ contractible.
More precisely, the topology change is described by a surgery procedure:
we cut out a tubular neighborhood of topology (3-ball)$\times {\rm R}^2\times {\rm S}^1$ from the ambient space $X$
and glue in a region of topology ${\rm S}^2 \times {\rm R}^2 \times$(disk) to get a new space $X'$.
A relative 2-cycle with boundary on $L$ combines with the disk
to become a 2-cycle in $X'$.

\begin{figure}[htbp]
\centering
\begin{tabular}{cccc}
\psfrag{M}{$m$}
\psfrag{L}{$l$}
\includegraphics[width=25mm]{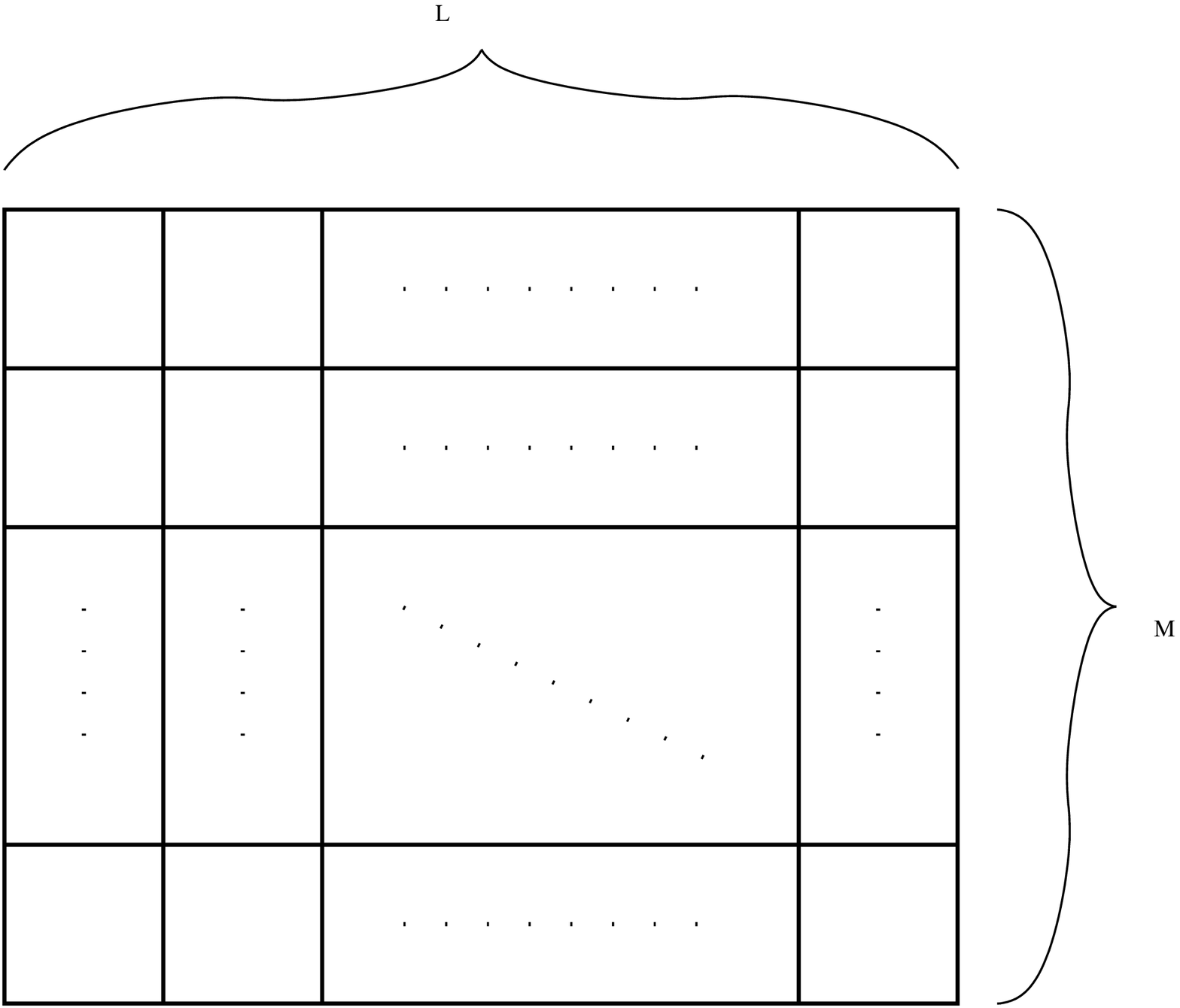}&
\psfrag{t1}{$t_1$}
\psfrag{t2}{$t_2$}
\psfrag{t3}{$t_3$}
\includegraphics[width=38mm]{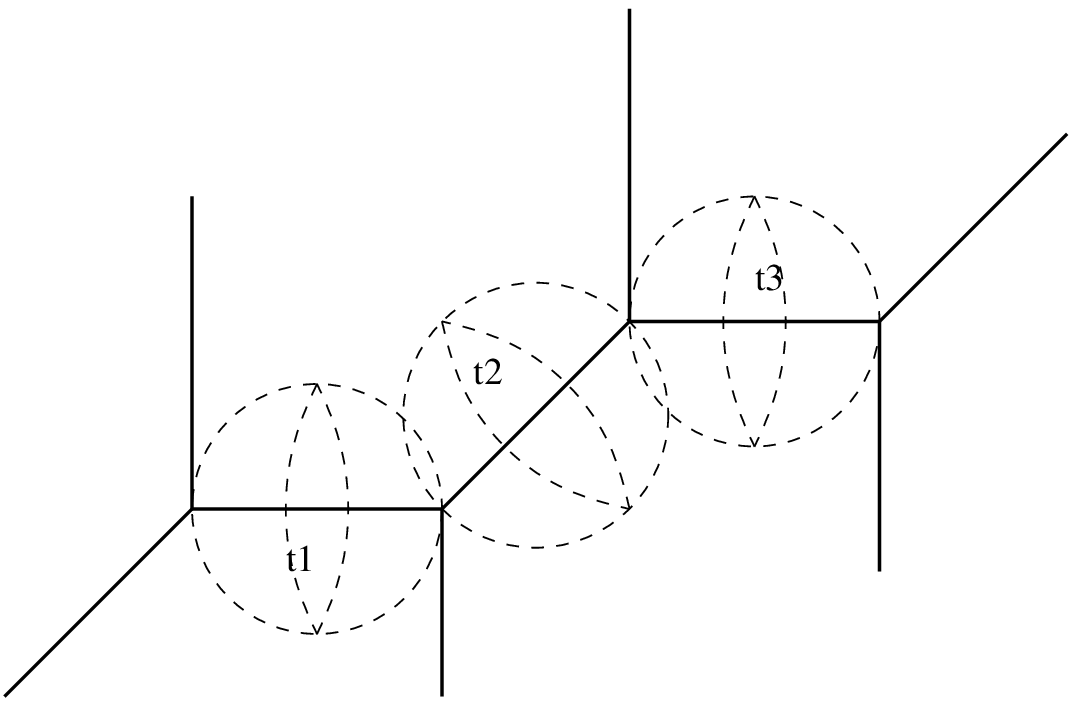}&
\psfrag{l1}{$\scriptscriptstyle{g_s(N+l)}$}
\includegraphics[width=38mm]{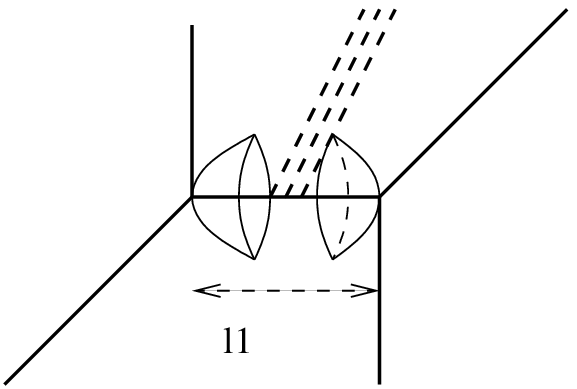}&
\psfrag{l2}{$\scriptscriptstyle{g_s(N-m)}$}
\includegraphics[width=38mm]{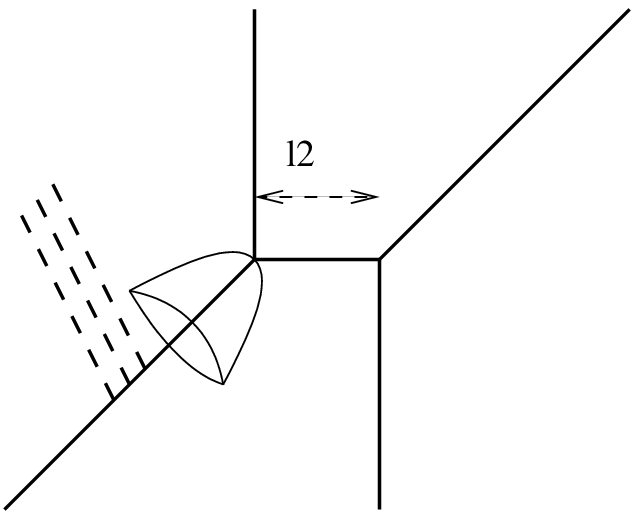}\\
(a)&(b)&(c)&(d)
\end{tabular}
\caption{
The simplest example of the correspondence between a Wilson loop and a bubbling
geometry.
The Wilson loop along the unknot in the $U(N)$ representation specified by the Young tableau
(a) is equivalent to the toric Calabi-Yau manifold given by the web
diagram (b).
The \Kahler moduli in (b) are given by $t_1=g_sm,~t_2= g_s l,~t_3=g_s
(N-m)$.
The bubbling CY manifold (b) arises from geometric transition of
$l$ branes in (c) as well as $m$ anti-branes in (d).
}
\label{bubbling-geometry_1}
\end{figure}

Let us now specialize to the Wilson loop along the unknot, in a representation
of the form Figure \ref{bubbling-geometry_1}(a).
The Wilson loop is realized by a stack of $l$ Lagrangian D-branes on the inner edge
at positions $x_i=g_s(l+m-i+1/2),~i=1,...,l$.
These branes live in the resolved conifold with \Kahler modulus
$t=g_s(N+l)$.
We propose that when $l$ and $m$ are large,
these branes undergo geometric transition to the toric Calabi-Yau
manifold whose toric web diagram is shown in
\ref{bubbling-geometry_1}(b).
The three \Kahler moduli are $t_1=g_s m,~t_2=g_s l, ~t_3=g_s(N-m)$.
Note that the new geometry has topology that is precisely as described above.
The contractible sphere that originally surrounded the branes become
a non-trivial cycle of size $g_s l$.
The two holomorphic disks that originally ended on the branes
become  spheres of sizes $g_s m$ and $g_s (N-m)$.
Our proposal is supported by the explicit computation for the unknot.
If we substitute $x_i=g_s(l+m-i+1/2)$ into the open$+$closed string partition function
(\ref{brane-amplitude})
in the brane set up of Figure \ref{bubbling-geometry_1}(c),
it becomes\footnote{For precise agreement we should include
$\xi(q)^{l}$ in (\ref{brane-amplitude}) where $\xi(q)=\prod_{j=1}^\infty (1-q^j)^\mo$.
This factor does not contribute to perturbative
amplitudes because of its modular property \cite{Saulina:2004da}.}
\ba
M(q)^2
\exp\sum_{n=1}^\infty \f{1}{n[n]^2}\left(
-e^{-n t_1}-e^{-n t_2}-e^{-n t_3}
+ e^{-n (t_1+t_2)} + e^{-n (t_2+t_3)}
-e^{-n (t_1+t_2+t_3)}\right)
\label{bubbling-amplitude_1}
\ea
This is precisely the closed string amplitude for the bubbling
Calabi-Yau in Figure \ref{bubbling-geometry_1}(b),
computed using mirror symmetry and integrality\footnote{See subsection 9.1 of \cite{Aganagic:2003db}
}!
This constitutes very strong evidence for our proposal for the geometric transition.

In section \ref{branes-res} we showed that the same Wilson loop is realized
by a stack of $m$ anti-branes on an outer edge with holonomies $y_i=g_s(l+m-i+1/2),~i=1,...,m$ turned on.
(See Figure \ref{bubbling-geometry_1}(d).)
The anti-branes live in the resolved conifold with $t=g_s(N-m)$.
These anti-branes also undergo geometric transition.
The resulting geometry is the same as for the branes on the inner edge.
In this case the 2-sphere surrounding the anti-branes acquires size $g_s m$,
and one 2-sphere of size $g_s l$ arises from a holomorphic disk ending on the anti-branes.
The open$+$closed string partition function
(\ref{anti-brane-amplitude})
for the anti-branes again becomes\footnote{
Insert $\xi(q)^{m}$ to (\ref{anti-brane-amplitude}) for precise agreement.
} the closed string partition
function (\ref{bubbling-amplitude_1}) after substituting the values for
$y_i$.

Let us now discuss the Wilson loop along the unknot in the general
representation $R$.
It is convenient to parameterize $R$ in terms of lengths $l_i$ of the edges
as in Figure \ref{bubbling-geometry_2}(a).
The open$+$closed string partition functions (\ref{brane-amplitude})
and (\ref{anti-brane-amplitude}) both become, after substituting the
values for holonomies,
\ba
&&M(q)^{m+1}\exp\sum_{n=1}^\infty \oo{n[n]^2}\left(
-\sum_{1\leq i\leq 2m+1} e^{-nt_i}
+\sum_{1\leq i\leq2m} e^{-n(t_i+t_{i+1})}\right.\nn\\
&&\left.~~
-\sum_{1\leq i\leq 2m-1} e^{-n(t_i+t_{i+1}+t_{i+2})}
...
- e^{-n(t_1+...+t_{2m+1})}
\right)\label{bubbling-amplitude_2}
\ea
We recognize this as  the closed string partition function for the toric
Calabi-Yau whose toric diagram is shown in Figure
\ref{bubbling-geometry_2}(b) \cite{Aganagic:2003db}!
This complicated geometry arises via
geometric transition  of branes or anti-branes.
The description in terms of geometry is more appropriate when
$l_i$, hence the cycles, are large. These Calabi-Yau manifolds provide  a novel representation of knot invariants in Chern-Simons theory.

In the forthcoming paper \cite{geom-trans} we will show that
the open GV invariants for an arbitrary knot determine the closed GV
invariants of the bubbling CY geometry as well as discuss generalization of the newly found class of geometric transitions.

\section*{Acknowledgments}
We thank Vincent Bouchard, Laurent Freidel, Amihay Hanany, Amir-Kian Kashani-Poor, Hirosi Ooguri, and Johannes Walcher for discussion.
We thank the Aspen Center for Physics where this project was
initiated.
The research of T.O. is supported in part by the NSF
grants PHY-9907949 and PHY-0456556.
Research of J.G. is supported in part by funds from NSERC of Canada and by MEDT of Ontario.

 \appendix

 \renewcommand{\theequation}{\Alph{section}\mbox{.}\arabic{equation}}

 \bigskip\bigskip
 \noindent {\LARGE \bf Appendix}


\section{Wilson loops as world-sheet instantons}\label{F1}

As in AdS/CFT,
Wilson loops in Chern-Simons theory are expected to be realized by non-compact
world-sheets \cite{Gopakumar:1998ki, Dymarsky:2006ve} in the resolved conifold.
In this appendix we explain how this works and point out that the description in terms of
world-sheets naturally gives rise to the generating functional of
Wilson loops originally obtained in \cite{Ooguri:1999bv}.
The generating functional is the amplitude for branes wrapping
a certain Lagrangian submanifold.
Conversely, this implies that exact computation of
the brane amplitude amounts to exact quantization of
string theory around a classical string solution.

In Chern-Simons perturbation theory, the expectation value of the Wilson loop
operator in the fundamental representation is computed by summing Feynman diagrams with external
legs ending on the loop:
\ba
\langle W_{\yng(1)}\rangle=\left\langle \Tr {\rm P} e^{\oint A}\right\rangle=
\begin{minipage}[h]{100mm}
\includegraphics[width=100mm]{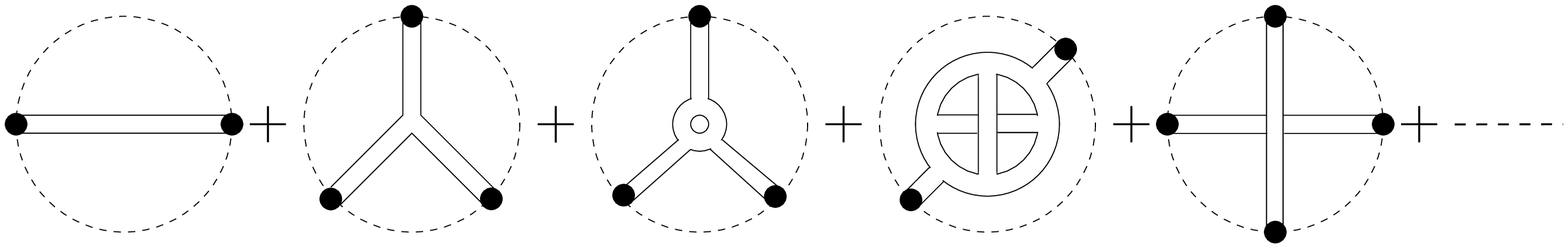}
\end{minipage}
\ea
Here the filled circles are the end points of the gauge field propagators, and
are integrated along the Wilson loop.
When the Feynman diagrams are drawn as fat graphs, they can
directly be viewed as infinitely thin open string world-sheets \cite{Witten:1992fb}.
To provide for a string theory description,
we place an extra holomorphic world-sheet that has the loop as its
boundary because the boundary of the world-sheet is the Wilson loop.
This surface is unique and is given as follows.
The equation for the deformed conifold is given in (\ref{def-con}).
Let us introduce $w$ coordinates by $z_1=w_1+iw_2,~z_4=w_1-i w_2,~z_2=w_3+iw_4,~z_3=-w_3+iw_4$.
Then the equation becomes
\ba
w_1{}^2+w_2{}^2+w_3{}^2+w_4{}^2=\mu,~~w_i\in \C,~i=1,2,3,4. \label{def-con_2}
\ea
Let us write $w_i=a_i+ i b_i$ with real $a_i, b_i$.
The $S^3$ is the real locus ($b_i=0$) of the hypersurface.
A knot can be parameterized as
\ba
a_i=f_i(\theta),~0\leq \theta\leq 2\pi,
\ea
where $f_i(\theta)$ are real functions satisfying $\sum_i f_i(\theta)^2=\mu$.
Then the holomorphic surface is given by
\ba
w_i(\zeta)=f_i(\zeta),~\zeta\in \C,~\zeta\sim \zeta+2\pi,~{\rm Im}\hspace{.5mm}\zeta\geq 0.
\label{eq-for-holo-cyl}
\ea
Thin open string world-sheets should then be glued to
the boundary of the holomorphic surface.
In other words, we are quantizing the open string theory around a
classical string solution represented by the holomorphic world-sheet.
As in AdS/CFT we need to cut off the infinite area of the surface.
After IR regularization the contribution of the infinite holomorphic surface is simply $e^{-{\rm Area}}$.
By subtracting the area, string theory around the holomorphic world-sheet computes the Wilson
loop vev.

It is easy to generalize this consideration to an arbitrary multi-trace
case
\ba
\langle W_{\vec k}\rangle:=
\left\langle \prod_{j=1}^\infty \left(\Tr {\rm P} e^{j\oint
A}\right)^{k_j} \right\rangle.
\ea
$\vec k$ is an infinite vector of non-negative integers.
If we know these expectation values we can compute the vev of the
Wilson loop in an arbitrary representation with the help of the
Frobenius relation.
The computation of such an operator is done by expanding the string
theory around a classical string solution in which we place $k_1$
singly wrapped surfaces, $k_2$ doubly wrapped surfaces, etc.
The classical configuration involves a total of $\sum k_j$ distinct
world-sheets.
Note that there are quantum fluctuations that connect them.
They represent the interactions among fundamental strings.

The IR regularization of the world-sheet can be conveniently done by introducing $M$ non-compact
D-branes on which the second boundary of the large world-sheet ends.

These D-branes wrap a Lagrangian $L$. It is in fact convenient to
deal at once with configurations with all possible $\vec k$ (with
$k_j=0$ for all $j>M$). Let $s$ be the area of the holomorphic
surface. The idea is that the surface appears in the perturbative
quantization of topological strings in the background with
non-compact D-branes. Let us denote by $V$ the $M\times M$ holonomy
matrix for the gauge field on $L$ along the boundary of the surface.
Because the world-volume is non-compact, the gauge field on the
non-compact branes is essentially non-dynamical. The contribution
from the stack of world-sheet instantons specified by $\vec k$ picks
up the factors  $\prod_{j=1}^\infty (\Tr V^j)^{k_j}$ from the
holonomy on the second boundary. The open string partition function
$Z(V)$ of the system with branes is given by \ba Z(V)=\sum_{\vec k}
\oo{z_{\vec k}}\langle W_{\vec k} \rangle\prod_{j=1}^\infty (\Tr
(e^{-s} V)^j)^{k_j}.\label{ZV} \ea $1/z_{\vec k}=1/\prod_j (k_j!)
j^{k_j}$ is the symmetry factor. Up to $e^{-s}$ this agrees with the
generating functional of Wilson loops obtained in
\cite{Ooguri:1999bv} through consideration in the open string
channel. Here we provided a closed string channel derivation.

Let us now follow the conifold transition to the resolved side.
The Gopakumar-Vafa duality implies that $Z(V)$ can be identified
with the partition function of topological strings in the
resolved conifold, in the presence of non-compact branes with
holonomy $V$.
The contribution to the coefficient of $\prod_{j=1}^\infty (\Tr
(e^{-s} V)^j)^{k_j}$ now comes from a stack of world-sheet
instantons.
Their boundaries are on $L$ and they are wrapped
multiply in the way specified by $\vec k$.
The branes provide IR regularization.
Thus our conclusion is that the Wilson loop $\prod_{j=1}^\infty \left(\Tr {\rm P} e^{j\oint
A}\right)^{k_j}$ is identified with the stack of non-compact world-sheet
instantons whose wrapping numbers are $\vec k=(k_1,k_2,...)$.
The exact computation of the brane amplitude (\ref{ZV}), which is possible for unknot, then
leads to exact quantization of string theory around the classical
string solution.
Note that the interactions between world-sheets, corresponding to
quantum fluctuations that connect world-sheets, arise in the perturbative
computation of the brane amplitude.

\section{Wilson loops as crystals} \label{crystal-section}
\label{section-crystal} In this section we revisit the crystal
description of the Wilson loop vev for the unknot, developed in
\cite{Okuda:2004mb}. Besides being interesting on its own right,
our aim here is to use  such crystals to facilitate the computations
showing that the Wilson loop vev equals the A-model amplitudes
of  branes, anti-branes, and the bubbling
Calabi-Yau's\footnote{The computations can in principle be performed
starting with the expression $\langle W_R\rangle\sim \det_{1\leq
i,j\leq N}(q^{j(R_i-i)})$, though the crystals provide more
transparent methods.}.

\begin{figure}[htbb]
\centering
\begin{tabular}{cccc}
\psfrag{l1}{$\scriptscriptstyle{l_1}$}
\psfrag{l2}{$\scriptscriptstyle{l_2}$}
\psfrag{l3}{$\scriptscriptstyle{l_3}$}
\psfrag{l4}{$\scriptscriptstyle{l_4}$}
\psfrag{l2m}{$\scriptscriptstyle{l_{2m}}$}
\psfrag{l2m+1}{$\scriptscriptstyle{l_{2m+1}}$}
\psfrag{N}{$N$}
\includegraphics[width=70mm]{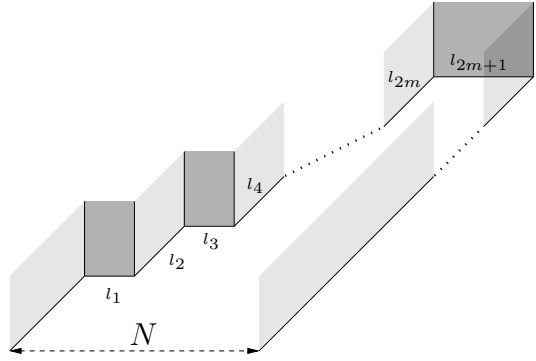}
\end{tabular}
\caption{Crystal configuration before melting.  This crystal corresponds to Figure \ref{bubbling-geometry_2}.}
\label{crystal_1}
\end{figure}

Let $R$ be the representation corresponding to a Young diagram.
Consider the three dimensional region shown in Figure
\ref{crystal_1}. A unit cube in the region is considered to be an
atom. Outside the region is the air. We regard the region as the
initial crystal configuration, which is specified by Young diagram
$R$ in Figure \ref{bubbling-geometry_2}(a). An atom can melt away
only if three of its faces are already exposed to the air \cite{Okounkov:2003sp}. Now let
us consider the ensemble of all crystal configurations in which some
atoms have melted away. The partition function is defined by summing
over all melted crystal configurations with Boltzman weight $q^E$,
where $E=(\hbox{number of removed boxes})$. It was shown in
\cite{Okuda:2004mb} that the unnormalized  expectation value of the
Wilson loop $\langle W_R\rangle$ for unknot agrees with the crystal
partition function up to some factors\footnote{ For the prefactors
that play no role for us, see \cite{Okuda:2004mb}.}.

This type of crystal melting problem can be treated efficiently
using the free CFT techniques \cite{Okounkov:2003sp}.
For our purposes, all the reader needs to know is the commutation relation of
two operators $\Gamma_\pm (z)$ and how they act on the vacuum\footnote{
If the reader wishes to understand how the crystal melting
problem is translated to the free CFT expressions,  formulas like $\Gamma_+(1)|R\rangle=\sum_{R\prec Q}|Q\rangle$ that can be
 found in \cite{Okounkov:2003sp}.}:
\ba
\Gamma_+(z_+)\Gamma_-(z_-)=\oo{1-z_+/z_-}\Gamma_-(z_-)\Gamma_+(z_+),~~
\Gamma_+(z)|0\rangle=\langle 0|\Gamma_-(z)=0. \label{Gamma+-}
\ea

The partition function $\langle W_R\rangle$ can be expressed in
several different ways.  Here we  consider the expressions which
differ only\footnote{ Our considerations here are all in the closed
string slicing as defined in \cite{Okuda:2004mb}. There is also the
open string slicing that is useful for relating $\langle W_R\rangle$
to the Wilson loop vev. } in the way the Young diagram $R$ is
parametrized. The first parametrization uses the number $R_i$ of boxes in
the $i$-th row. The second uses the number $R^T_i$ of boxes in the
$i$-th column. The third uses the length $l_i$ of $i$-the edge. See
Figures \ref{young-param-1} and \ref{bubbling-geometry_2}(a) for  examples.

In the parametrization in terms of the
number $R_i$ of boxes in the $i$-th row, the Wilson loop vev is given by
\ba
\langle W_R\rangle&=&
\langle 0|\left[ \prod_{i=R_1}^\infty\Gamma_+(q^i)\right]
\Gamma_-(q^{R_1-1})
\left[ \prod_{i=R_2-1}^{R_1-2}\Gamma_+(q^i)\right]
\Gamma_-(q^{R_2-2})...\nn\\
&&\times
\Gamma_-(q^{R_P-P})
\left[\prod_{i=-P}^{R_P-P-1}\Gamma_+(q^i)\right]
\left[\prod_{i=-N}^{-P-1}\Gamma_-(q^i)\right]
|0\rangle. \label{crystal-R}
\ea
(\ref{Gamma+-}) allows one to compute this and obtain
(\ref{anti-brane-amplitude}) that is the partition function in the presence
of anti-branes on the outer edge.
The special case $P=1$ of this interpretation was found in \cite{Halmagyi:2005vk}.

The parametrization in terms of $R^T_i$, the number of boxes in each column,
naturally relates the
crystal to the branes in section \ref{branes-res}.
This is the parametrization considered in \cite{Okuda:2004mb}.
It was found that\footnote{
The expressions in (\ref{crystal-R}-\ref{crystal-l}) differ in
the parametrization of the Young diagram as well as the overall rescaling
in the arguments of $\Gamma_\pm$.
Neither affects the amplitude.
The overall rescaling is equivalent to conjugating all the operators by a power of $q^{L_0}$,
where $L_0$ is the Virasoro zero mode \cite{Okounkov:2003sp}.
}
\ba
\langle W_R\rangle
&=&\langle 0|\left[\prod_{i=1}^\infty \Gamma_+(q^{i-1})\right]
\left[\prod_{i=1}^{R^T_M}\Gamma_-(q^{-i})\right]\Gamma_+(q^{-(R^T_M+1)})
\left[\prod_{i=R^T_M+2}^{R^T_{M-1}+1}\Gamma_-(q^{-i})\right]...\nn\\
&&\times
\left[\prod_{i=R^T_2+M}^{R^T_1+M-1}
\Gamma_-(q^{-i})\right]\Gamma_+(q^{-(R^T_1+M)})
\left[\prod_{i=R^T_1+M+1}^{N+M}\Gamma_-(q^{-i})\right]|0\rangle.
\label{crystal-RT}
\ea
One can derive
(\ref{brane-amplitude}) from this.

The parametrization in terms of $l_i$ is useful for relating the
crystal to the bubbling geometry.
$\langle W_R\rangle$
is written as
\ba
&&\langle W_R\rangle=\langle 0|
\left[ \prod_{i=1}^\infty \Gamma_+(q^{i-1})\right]
\left[ \prod_{i=1}^{l_1} \Gamma_-(q^{-i})\right]
\left[ \prod_{i=1}^{l_2} \Gamma_+(q^{-l_1-i})\right]...\nn\\
&&\times
\left[ \prod_{i=1}^{l_{2m-1}} \Gamma_-(q^{-l_1-l_2-...-l_{2m-2}-i})\right]
\left[ \prod_{i=1}^{l_{2m}}
\Gamma_+(q^{-l_1-l_2-...-l_{2m-1}-i})\right]\nn\\
&&\times
\left[\prod_{i=1}^{l_{2m+1}}
\Gamma_-(q^{-l_1-l_2-...--l_{2m-1}-l_{2m}-i})\right]|0\rangle. \label{crystal-l}
\ea
Using (\ref{Gamma+-}), one can show that this reduces to (\ref{bubbling-amplitude_2}).
The crystal is  directly related to the bubbling
geometry as one sees by comparing Figures \ref{bubbling-geometry_2} and \ref{crystal_1}.

\section{Target space theory of open topological strings} \label{target-theory}

Let us consider the case of a single A-brane wrapping a Lagrangian
submanifold $L$ of a Calabi-Yau manifold $M$.
In A-model, only (the cohomology class of) the symplectic structure
$\omega=\half \omega_{\mu\nu}dx^\mu\wedge dx^\nu$ of $M$, not the whole \Kahler structure, is relevant.
When the B-field vanishes, the equation of motion for the gauge
field is simply $F\equiv dA=0$.
When the B-field is turned on the equation of motion should be
generalized so that it is invariant under the gauge transformation
$B\ra B+d\Lambda,~A\ra A-\Lambda$\footnote{The notation for gauge
fields in this appendix differs from the rest of the paper.  Here
$A$ is a hermitian connection, and $\Acal=\phi-iA$ is its complexification.
}.
The minimal
gauge invariant equation of motion is
\ba
F+B=0.
\ea

For a single Lagrangian A-brane, one real modulus is given, when
$B=0$, by the holonomy $\oint A$ along a non-trivial 1-cycle of
$L$.  This is again to be promoted to a gauge-invariant combination.
The minimal combination is
\ba
\oint_{\p D}A +\int_D B, \label{A-B-modulus}
\ea
where $D\subset M$ is a disk (or more generally a surface with boundary) whose boundary is the
1-cycle\footnote{
Since $b_1(M)=0$, the 1-cycle of $L$ has to be the boundary of
some chain $D$ in $M$.
}.

In topological field theory, all moduli are complex.
There must be another real modulus
associated with the 1-cycle that combines with (\ref{A-B-modulus})
to form a complex modulus.
The existence of the additional modulus can be understood as
follows \cite{Aspinwall:2004jr}.

Let us consider an infinitesimal
deformation of $L$ that preserves the condition that the submanifold is Lagrangian.
An infinitesimal deformation corresponds to a
normal vector field $v$, i.e., a section of the normal bundle
$TM|_L/TL$.  We can think of $v$ as a vector field defined in a
neighborhood of $L$
\ba
v=v^\mu\p_\mu
\ea
(indices are for $M$ coordinates), up to a vector field $\xi$,
defined in the same domain, such that $\xi|_L\in TL$.
$\omega$ is preserved by the deformation if and only if
\ba
L_v \omega=0
\ea
where $L_v$ is the Lie derivative.
By contracting
$v$ with the symplectic form $\omega$,
one can form a 1-form in the neighborhood of $L$:
\ba
i_v\omega=\omega_{\mu\nu}v^\mu dx^\nu.
\ea
$i_v \omega$ is closed
and its restriction $ i_v\omega|_L$ defines
a cohomology class of $L$.
$i_v\omega|_L$ is exact if and only if $v$ is a Hamiltonian vector field:
$v=(\omega^\mo)^{\mu\nu}\p_\nu f$ for some function $f$.
Indeed it is known that two Lagrangian submanifolds related by
a Hamiltonian deformation defines the same open string theory \cite{Aspinwall:2004jr}.
Infinitesimally, the deformation modulus is given by $\oint_{\p
D}i_v\omega$.

The modulus for a finite deformation is known to be given (for a single brane)
by the symplectic area of the disk\footnote{
When $D$ is a holomorphically embedded disk, the symplectic area
agrees with the area with respect to the \Kahler metric of $M$.
}
\ba
\int_D\omega. \label{omega-modulus}
\ea
One can easily check that the variation of (\ref{omega-modulus}) is
proportional to $\oint_{\p
D}i_v\omega$.

The description of the deformation modulus in the above two paragraphs is rather unsatisfactory
because unlike (\ref{A-B-modulus}),
 (\ref{omega-modulus}) does not depend on a field of the
world-volume theory. We now point out that the correct description
of the brane, namely the Chern-Simons theory with a complex gauge
group, generalizes (\ref{omega-modulus}) into a form that is in
parallel with (\ref{A-B-modulus}).

In string theory, all deformations of the theory must appear in the
physical spectrum.  As Witten showed in \cite{Witten:1992fb},
the spectrum comes solely from $H^1(L)$.
For $N$ coincident branes, this gives rise to
$U(N)$ theory when a reality condition is imposed on the
string field.  To account for deformations of the Lagrangian submanifold,
however, it is more natural not to impose the
reality condition \cite{Aspinwall:2004jr}.
Thus the gauge field is a complex field, and the
gauge group is $GL(N,{\rm C})$.
We write $A+i\phi$ for the complex gauge field.

To understand the meaning of the imaginary part of the gauge field,
let us consider the vertex operators for the gauge fields.
Let us use coordinates $X^a$ to parameterize $L$, and $Y^a$ for transverse
directions so that $X^a+iY^a$ are complex coordinates of $M$.
Topological string theory is a version of bosonic string theory
\cite{Witten:1992fb}.
The ``fixed'' vertex operators \cite{Polchinski:1998rq} for the real gauge fields $A_a$ are the
fermions $\chi^a$ in the notation of \cite{Witten:1992fb}.
The corresponding ``integrated'' vertex operators are
$\{b_0,\chi^a\}$, which turn out to be $\p_\tau X^a$.
($\tau$ is the parameter for the boundary of a world-sheet, and $X^{a}(z=\sigma+i\tau)$ are
the embedding coordinates.)
Thus the ``integrated'' vertex operators for the imaginary gauge
fields $\phi_a$ are $i\p_t X^a$.
In topological A-model, only the field configurations such that
$X^a+iY^a$ is holomorphic in $z$ contribute.
For such configurations $i\p_\tau X^a=-i\p_\sigma Y^a$.
It is well-known that integrating these vertex operators induces
transverse deformations of the brane.
Thus we have shown that the imaginary gauge fields give rise to
deformations of the brane.

Our conclusion, which is in agreement with some of the literature
\cite{Aspinwall:2004jr}, is that in topological A-model the natural
gauge group for the Chern-Simons theory on A-branes is a complex
gauge group.  This obviously raises the issue of non-perturbative
definition because the path-integral seems to be divergent.  Our
attitude here is that only the perturbative definition of the Chern-Simons theory
is relevant.  This is consistent with the fact that topological
string theory is defined only perturbatively.

Once we accept the complex gauge group,
there is natural generalization of (\ref{omega-modulus})
for the real modulus:
\ba
\oint_{\p D}\phi+\int_D\omega
\ea
In other words, the complex modulus corresponding to the 1-cycle is
\ba
\oint_{\p D}\Acal+\int_D \Jcal,
\ea
where $\Acal\equiv \phi-iA,~~\Jcal \equiv (\omega-iB)$.
Note that this expression is invariant under the complex generalization of the B-field gauge transformation
\ba
B+i\omega\ra B+i\omega+d(\Lambda+i\lambda),~~A+i\phi\ra
A+i\phi- (\Lambda+i\lambda).
\label{J-gauge-transform}
\ea

The minimal non-Abelian gauge invariant equation of motion
for the open strings interacting with closed strings is
\ba
\Fcal+\Jcal 1_{N\times N}=0,
\ea
where $\Fcal=d\Acal+\Acal^2$.

For a promising approach to the target space theory,
see \cite{Thomas:2001ve}.

\bibliography{loop}

\end{document}